\documentclass[a4paper,11pt,superscriptaddress]{article}
\usepackage{jheppub} 
\usepackage{braket}

\usepackage{tikz-feynman}
\usetikzlibrary{calc,decorations.pathmorphing}
\tikzfeynmanset{compat=1.1.0}
\usepackage{lineno}
\usepackage{hyperref}
\usepackage[capitalise]{cleveref}
\usepackage{booktabs}
\usepackage{longtable}
\usepackage{tabularx}
\usepackage{array}
\usepackage{float}
\usepackage{makecell}

\newcommand{\sns}{Scuola Normale Superiore and INFN, Piazza dei Cavalieri 7, 56126, Pisa, Italy}
\newcommand{\hu}{Jefferson Physical Laboratory, Harvard University, Cambridge, MA 02138, USA}

\title{\boldmath Local Measures in Odd de Sitter}

\begin{document}

\author[a]{Guilherme L. Pimentel,}
\author[a,b]{Nicole Righi,}
\author[a]{and Behrang Tafreshi}
\affiliation[a]{\sns}
\affiliation[b]{\hu}

\abstract{We show that massless theories in odd-dimensional de Sitter space have very simple measures---the modulus squared of the late-time wavefunctional---at tree level. They are local functionals; in particular, all wavefunction coefficients vanish at separated positions. All the usual singularities of nonlocal wavefunctions in even dimensions only affect the phase in the odd dimensional case. Hence a closed formula can be written for any number of points. We focus on three examples: the non-linear sigma model, Yang-Mills theory and General Relativity, from dimensions three to seven.}

\maketitle
\flushbottom
\setcounter{page}{1}
\section{Introduction}
The probability measure for massless field fluctuations in de Sitter space, given by the absolute value squared of the vacuum wavefunction, is typically a complicated functional, with wavefunction coefficients having support at separated field insertions. The wavefunction coefficients are the closest analogues of scattering amplitudes in cosmology,\footnote{Though see \cite{Marolf:2012kh,Melville:2023kgd,Melville:2024ove,Taylor:2026ase} for proposals of massive de Sitter S-matrices.} and a systematic development of this analogy has been very fruitful in recent years (see e.g.~\cite{Benincasa:2022gtd} and references therein). Due to redshift, conjugate momenta tend to have vanishingly small amplitude, meaning that the de Sitter vacuum is a heavily squeezed state. For that reason, typically we are interested in computing only the absolute value squared of the wavefunction, or equivalently the real part of wavefunction coefficients. The probability measure, i.e.~the {\it real} part of the wavefunction coefficients, is the object from which we can compute quantum expectation values of fields at late times~\cite{Maldacena:2002vr}.  In this paper we point out that massless interacting gauge theories in {\it odd-dimensional} de Sitter space have local probability measures,\footnote{Of course, measurements of correlations between the late-time distribution of the field and its conjugate momentum would be sensitive to the phase of the wavefunction, which is nonlocal.} which we write explicitly. This implies that they are analytic functions of the momenta.

In four-dimensional gauge theories, the organization of amplitudes on helicity sectors has proven to be particularly useful. At generic kinematics, tree-level all-plus and single-minus Yang-Mills amplitudes vanish, the first essentially because of supersymmetry, the second due to classical integrability of the self-dual sector of Yang-Mills theory \footnote{For recent studies of self-dual Yang--Mills theory and extensions, see \cite{Skvortsov:2026dru,Heuveline:2026knw, Chowdhury:2026jpt,Gomez:2026wqt}.} (similar statements apply to gravity as well). However, the MHV sector, described at arbitrary multiplicity by the Parke-Taylor formula~\cite{Parke:1986gb}, is non-zero at very special kinematics. This has been highlighted in a series of recent papers~\cite{Monteiro:2011pc,Guevara:2026qzd,Guevara:2026qwa} which have moreover found a general form of the amplitude at $n$ points. The same phenomenon persists in gravity. Finally the MHV amplitude is meromorphic and has support for any value of the momenta. Of course, all of these statements hold at tree-level.\footnote{At one-loop, the all-plus amplitude is non-zero but can still be written in a very simple form, being a rational function of the momenta~\cite{Bern:1994zx}.} 

Local measures like the ones presented in this work provide, in our opinion, the closest analogue to special amplitudes. This is because the measure is only nonzero when all field insertions at the boundary are at the same location. In other words, the singularities associated to propagators are absent, and yet there is a nontrivial measure. Despite the locality of the measure, the expectation values of fields are nonlocal, although simpler than one would naively expect.\footnote{The simplest example is the quadratic coefficient of the measure. For example, in gravity it goes as $k^d$ in $d+1$ dimensions, which is analytic. Yet the expectation value of the metric goes as $k^{-d}$, which is nonlocal and characteristic of scale invariance in inflation.} 

We make this quantitative in \cref{sec:setup}. We outline the setup in full generality, and explain why the finite nonlocal contributions to the wavefunction enter only its phase in odd dimensions. This is followed by reviewing one example of this phenomenon, first shown in \cite{Maldacena:2011mk}, namely 5d Einstein gravity in de Sitter (dS) space. There, the data of the late-time probability measure is a Euclidean four-metric in the boundary, which is taken to have flat topology, and reads
\begin{equation}
    |\Psi_{\rm dS}[\hat{g}]|^2=\exp\left(-\frac{M_{\rm pl}^3R_{\rm dS}^3}{2}\frac{\pi}{8}\int \sqrt{\hat{g}}\, \hat W[\hat{g}]^2\right)\, ,
    \label{eq:gravity_wavefunction_5d}
\end{equation}
where $\hat W$ is the Weyl tensor of the boundary metric $\hat g$. This is a remarkably simple contribution, predicting the tree-level measure at any number of points, while in e.g.~4d gravity in de Sitter we have only recently been able to compute its analogue beyond four-points! Of course in such a theory cosmological correlators are nontrivial, and computing them remains a difficult task. However, they all contribute to the phase of the wavefunction, leaving no imprint on the measure. The fact that this contribution stems from infrared divergences of the bulk theory, explains locality through the holographic principle \cite{Susskind:1998dq}.

We further push this observation and show that Einstein gravity, Yang-Mills theory, and non-linear sigma models (NLSM) in odd-dimensional de Sitter space have similarly simple, local measures. The results are summarized in \cref{tab:summary-gr-YM-NLSM}. Besides deriving the measures, in \cref{sec:direct_check_and_origin} we report checks of our formulae from perturbation theory and the cosmological CPT theorem \cite{Goodhew:2024eup}. Finally, in \cref{sec:conclusion}, we comment on possible generalizations to semi-local measures, special cases in three dimensions, and potential avenues for a deeper understanding of the origin of locality. Since the present work is restricted to tree level, we also comment on possible one-loop contributions, leaving a full analysis for future work.

\section{EAdS Origin of the Local Measure}
\label{sec:setup}

The primary objective of this work is to determine the probability measure of the late-time quantum wavefunction in de Sitter space for a broad class of theories, namely gravity, Yang-Mills theory, and non-linear sigma models. Computing the cosmological wavefunction is generally a technically demanding problem: each coefficient receives contributions from contact and exchange diagrams involving bulk time integrals, together with boundary counterterms required to regulate late-time divergences. The result typically has a complicated nonlocal dependence on the external momenta. In the odd-dimensional theories considered here, however, it will turn out that all of this complexity is confined to the phase of the wavefunction, while the probability measure is determined entirely by a local logarithmic contribution.  To see this, we spell out in detail the analytic continuation from Euclidean Anti-de Sitter (EAdS) and its features.

\subsection{The EAdS argument}
We consider a quantum field in $(d+1)$-dimensional EAdS, using the Fefferman-Graham coordinate system
\begin{equation}
    ds^2_{\rm EAdS}
    = G_{\mu\nu} dx^\mu dx^\nu = 
    \frac{R_{\rm AdS}^2}{z^2}
    \left(
        dz^2+ g_{ij}(z,x) d x^i d x^j
    \right)\,,
    \quad z>0\,,
    \label{eq:metricFG}
\end{equation}
with $i, j = 1, \dots, d$. Here $g_{ij}(z,x)$ denotes the bulk spatial metric that reduces to the boundary metric $\hat g_{ij}(x)$ at $z \to 0$.

The late-time de Sitter wavefunction $\Psi_{\rm dS}$ is related to the EAdS partition function $Z_{\rm AdS}$ via the double analytic continuation 
\begin{equation}\label{eq:double_analytic_continuation}
    R_{\rm AdS}\rightarrow iR_{\rm dS}\,,
    \qquad
    z\rightarrow -i\eta\,,
    \qquad
    \eta<0\,, 
\end{equation}
which maps the EAdS metric to the flat-slicing Lorentzian de Sitter metric:
\begin{equation}
    ds^2_{\rm dS}
    =
    \frac{R_{\rm dS}^2}{\eta^2}
    \left(
        -d\eta^2+g_{ij}(\eta,x) d x^i d x^j
    \right)\,.
\end{equation}
At the semiclassical level, the Euclidean partition function is given by the exponential of the on-shell action as $Z_{\rm AdS} \simeq \exp\left(-S^{\rm EAdS}_{\rm os}\right)$, and its analytic continuation yields the Lorentzian de Sitter wavefunction
\begin{equation}
    Z_{\rm AdS} \simeq \exp\left(-S^{\rm EAdS}_{\rm os}[\hat{A}]\right)\;
    \longrightarrow\;
    \Psi_{\rm dS}[\hat{A}] \simeq \exp\left(iS^{\rm dS}_{\rm os}[\hat{A}]\right)\,,
\end{equation}
where $\hat{A}$ stands for the Dirichlet boundary condition imposed for the field $A$. Note that unlike $S_{\rm os}^{\rm EAdS}$ which is strictly real, $S_{\rm os}^{\rm dS}$ is complex.

Our main focus is on odd-dimensional de Sitter spaces, corresponding to even-dimensional conformal boundaries. For the theories considered here, the bulk actions carry an odd power of the AdS radius:
\begin{equation}
    S_{\rm grav},\,S_{\rm NLSM} = R_{\rm AdS}^{d-1} \times (\cdots)\,,
    \qquad
    S_{\rm YM} = R_{\rm AdS}^{d-3} \times (\cdots)\,,
\end{equation}
where both exponents are odd when the bulk dimension $(d+1)$ is odd. Under analytic continuation $(R_{\rm AdS})^{\text{odd}} \to i(\pm R_{\rm dS})^{\text{odd}}$, meaning that real terms in the Euclidean on-shell action are in general mapped to phases of the Lorentzian wavefunction.

To compute the late-time probability measure,
\begin{equation}
    \log|\Psi_{\rm dS}|^2
    =
    -2\,\operatorname{Im}S_{\rm os}^{\rm dS}\,,
\end{equation}
we follow similar steps as in \cite{deHaro:2000vlm,Skenderis:2002wp}. The regulated EAdS on-shell action takes the schematic form 
\begin{equation}
    S^{\rm EAdS}_{\rm os}
    =
    R_{\rm AdS}^{\,p}
    \left[
        \sum_n \frac{S^{\rm EAdS}_{(2n)}}{\epsilon^{\,2n}}
        +
        S_{\log}^{\rm EAdS}\log\epsilon
        +
        S_{\rm fin}^{\rm EAdS}
    \right]\,,
    \qquad p\ {\rm odd}\,,
    \label{eq:reg-onshell-action-form}
\end{equation}
where $\epsilon$ is the radial IR cutoff. Importantly, this structure relies on the analyticity of the bulk on-shell action in $z$, as is the case for gravity, Yang-Mills theory, and the NLSM considered here. For a scalar field of generic mass, however, this analyticity is generally lost, and the on-shell action need not admit an expansion of the form above. Focusing on the analytic case described by \cref{eq:reg-onshell-action-form}, we now continue to de Sitter, under which the radial cutoff rotates as $\epsilon=-i\epsilon_\eta$ ($\epsilon_\eta<0$), yielding
\begin{equation}\label{eq:monodromy}
    \log\epsilon
    \longrightarrow
    \log(-\epsilon_\eta)+\frac{i\pi}{2}\,.
\end{equation}
Crucially, while power-law divergences and finite real terms only contribute to the phase (i.e., the imaginary part of the exponent) of the wavefunction, the logarithmic term produces a real contribution due to the $i\pi/2$ shift, namely
\begin{equation}
    \begin{aligned}
        S_{\rm os}^{\rm EAdS} \rightarrow i S_{\rm os}^{\rm dS}  &= (-1)^{\frac{p+1}{2}} \, i R_{\rm dS}^p \left[
        \sum_n (-1)^n \frac{S^{\rm dS}_{(2n)}}{\epsilon_{\eta}^{\,2n}}
        +
        S_{\log}^{\rm dS}\log(-\epsilon_{\eta})
        +
        S_{\rm fin}^{\rm dS}
        \right] \\ 
        & \qquad  + (-1)^{\frac{p+3}{2}} \frac{\pi}{2} R_{\rm dS}^p S_{\log}^{\rm dS}\,.
    \end{aligned}
\end{equation}
Consequently, the probability measure is controlled entirely by the coefficient of the logarithmic divergence:
\begin{equation}
    \log |\Psi_{\rm dS}|^2
    \;=\; (-1)^{\frac{p+3}{2}}
    \pi\, R_{\rm dS}^p S_{\log}^{\rm dS}\,.
    \label{eq:log-measure}
\end{equation}
Thus, the problem of determining the probability measure reduces to computing the regulated EAdS on-shell action for each theory, isolating its logarithmic divergence $S_{\log}^{\rm EAdS}$, and continuing it to de Sitter.

Now, we shall illustrate that the measure for odd-dimensional de Sitter space is local for gravity, Yang-Mills theory, and NLSM. An important distinction between these theories concerns the role of the boundary metric, which is itself dynamical in the gravitational case. For gravity, we retain the general metric $\hat g_{ij}$ at the boundary, whereas for Yang-Mills and NLSM we restrict to a flat boundary for simplicity.\footnote{For a curved boundary, one can find the result following the same procedure presented here.}

\subsection{Einstein gravity}
We first recall the general argument applied to $5$d Einstein gravity and subsequently state the result of the $7$d case. The resulting probability measures, corresponding to 4d and 6d conformal gravity, are summarized in \cref{tab:summary-gr-YM-NLSM}.

The setup is as follows. We use Dirichlet boundary condition to fix $\hat{g}$ as the boundary metric. We also have a bulk Einstein metric obeying the usual $d+1$-dimensional Einstein manifold condition, $R_{\mu\nu} = -d g_{\mu\nu}$. The regularized on-shell action of \cref{eq:reg-onshell-action-form} for pure gravity takes the form~\cite{Henningson:1998gx}\footnote{We assume the boundary topology to be trivial. A nontrivial boundary topology can also be accounted for by the addition of a Gauss-Bonnet term.}
\begin{equation}
    \begin{aligned}
    -S[\hat{g}] &= \frac{M_{\rm pl}^{3} R_{\rm AdS}^{3}}{2}  \left[ \int_{z\geq \epsilon} d^{5}x \sqrt{G} (R + 12) + 2 \int d^{4}x K \right] \\
    & = \frac{M_{\rm pl}^{3} R_{\rm AdS}^{3}}{2} \Big[ \frac{1}{\epsilon^4} \int d^{d}x \sqrt{\hat{g}} \,{a_0} + \frac{1}{\epsilon^{2}} \int d^{d}x \sqrt{\hat{g}}\, a_2  - \frac{\log \epsilon}{8}\int \sqrt{\hat{g}} \,  \hat{W}^2  - S_{\rm fin}[\hat{g}]\,\Big]\,.
    \end{aligned}
\end{equation}
Upon analytic continuation (\cref{eq:double_analytic_continuation}), it is only the logarithmic divergence that contributes to the measure of the theory in de Sitter space, which is why we only state such term explicitly. Substituting the above expansion into \cref{eq:log-measure} gives the result shown in \cref{tab:summary-gr-YM-NLSM}.

\subsection{Yang-Mills theory}
We next apply the same framework to Yang-Mills theory. We derive the late-time probability measure explicitly in $5$d de Sitter space and summarize its $7$d counterpart in \cref{tab:summary-gr-YM-NLSM}.

The action of Yang-Mills theory on an asymptotically EAdS$_5$ reads
\begin{equation}
    S_E^{\rm YM}
    =
    \frac{1}{4g_5^2}
    \int_{z\geq \epsilon} d^5x\,\sqrt{G}\,
    F^a_{\mu\nu}F^{a\,\mu\nu}\,,
    \label{eq:YM-action}
\end{equation}
where $F^a_{\mu\nu}=\partial_\mu A^a_\nu-\partial_\nu A^a_\mu +f^{abc}A^b_\mu A^c_\nu\,$. In five dimensions $g_5^2$ has dimensions of length, and hence $\tfrac{R_{\rm AdS}}{g_5^2}$ is dimensionless. We impose radial gauge $A_z=0$ and Dirichlet boundary conditions
\begin{equation}
   A_i(z,x) \xrightarrow[z \rightarrow 0]{} \hat{A}_i(x)\,.
    \label{eq:YM-Dirichlet}
\end{equation}
The residual $z$-independent gauge transformations act on $\hat A_i$ as
an ordinary 4d gauge transformation.  Denote the field strength of $\hat{A}_i$ as
\begin{equation}
    \hat F_{ij}= \partial_i\hat A_j-\partial_j\hat A_i
    +[\hat A_i,\hat A_j]\,,
\end{equation}
and restricting to a flat boundary, we use
\begin{equation}
    \hat D_i X=\hat\partial_i X+[\hat A_i,X]
\end{equation}
for the derivative covariant with respect to both the boundary metric
$\hat g_{ij}$ and the gauge connection $\hat A_i$. In radial gauge, the action can consequently be written as
\begin{equation}
    S_E^{\rm YM}
    =
    \frac{R_{\rm AdS}}{4g_5^2}
    \int_{z\geq \epsilon} \frac{dz}{z}
    \int d^4x\,\sqrt{g}\,
    \left[
        F^a_{ij}F^{a\,ij}
        +
        2g^{ij}\partial_zA^a_i\partial_zA^a_j
    \right]\,.
    \label{eq:YM-scaled-action}
\end{equation}
Thus the complete classical on-shell action carries the common dimensionless prefactor $\tfrac{R_{\rm AdS}}{g_5^2}$, which is the Yang-Mills analogue of the gravitational coefficient $(M_{\rm pl}L)^{3}$.

In order to obtain the near-boundary expansion of the gauge field, we perturbatively expand the metric and the gauge field in $z$ and solve the equation of motion order by order. The equations of motion read
\begin{equation}
    D_\mu \big(\sqrt{G}\,F^{\mu\nu}\big)=0\,.
\end{equation}
Imposing the Dirichlet boundary condition we get
\begin{equation}
    A_i(z,x) =\hat A_i(x)
    +z^2\left[
        A_{(2)i}(x) -\frac12\log(\mu z)\hat D^j\hat F_{ji}(x)
    \right] +\dots\, .
    \label{eq:YM-final-expansion}
\end{equation}
Here, the appearance of the logarithmic term is due to a resonance in the equation of motion. Note that the coefficient of the logarithmic term is purely local in the boundary data. By contrast, $A_{(2)i}$ is not completely fixed by the near-boundary analysis.  It is fully determined only after imposing an appropriate condition in the interior, such as regularity in EAdS, and therefore becomes in general a nonlocal functional of the
boundary source $\hat A_i$. Using \cref{eq:YM-final-expansion}, we see  that the second term in \cref{eq:YM-scaled-action} is IR integrable, whereas integrating the first term over the radial direction produces
\begin{equation}
\begin{split}
    S_{E,\epsilon}^{\rm YM}
    & = -\frac{R_{\rm AdS}}{4g_5^2} \log \epsilon
    \int d^4x \,
    \hat F^a_{ij}\hat F^{a\,ij}
    +S_{\rm fin}\,.
\end{split}
\end{equation}
We find that the coefficient of the logarithmically divergent part is the 4d Yang-Mills theory on the boundary. Consequently, after analytic continuation, the dS$_5$ wavefunction of Yang-Mills theory reads
\begin{equation}
    |\Psi_{\rm dS}[\hat{A}]|^2_{\rm YM} = \exp \left[ -\frac{ R_{\rm dS}}{4 g_5^2} \pi \int d^4 x \, \hat F^a_{ij}\hat F^{a\,ij} \right]\,.
    \label{eq:YM5_dS}
\end{equation}

\subsection{Non-Linear Sigma Model}
Finally, we turn to applying this framework to the non-linear sigma model. Let $\Phi^A(X)$ be the coordinates for a map from the bulk spacetime to a target manifold $\mathcal{M}$ with target metric $\mathcal G_{AB}(\Phi)$. The $5$d Euclidean NLSM action on the fixed EAdS background in \cref{eq:metricFG} reads
\begin{equation}\label{eq:actionNLSM}
    \begin{aligned}
        S_{\rm NLSM}
        =\frac{R_{\rm AdS}^3}{2g_5^2}\int_{z \geq \epsilon}\frac{dz}{z^3}\int d^4x\sqrt{g}\,\mathcal G_{AB} (\Phi)\left(\partial_z\Phi^A\partial_z\Phi^B+g^{ij}\partial_i\Phi^A\partial_j\Phi^B\right)\,.
    \end{aligned}
\end{equation}
The equations of motion are
\begin{align}
  \partial_z^2\Phi^A+\Gamma^A_{BC}(\Phi)
\partial_z\Phi^B\partial_z\Phi^C+\left(
\frac{\partial_z\sqrt g}{\sqrt{g}}-\frac3z
\right)\partial_z\Phi^A+\mathcal D^i\partial_i\Phi^A=0\,,  
\end{align}
where $\Gamma^A_{BC}(\Phi)$ is the Christoffel symbol with respect to $\mathcal{G}_{AB}(\Phi)$ and we have introduced the covariant derivative
\begin{equation}
    \mathcal D^i\partial_i\Phi^A=\frac1{\sqrt g}\partial_i\left(\sqrt g\,g^{ij}\partial_j\Phi^A\right)+\Gamma^A_{BC}(\Phi)
g^{ij}\partial_i\Phi^B\partial_j\Phi^C\,.
\end{equation}
To solve them, we expand $\Phi^A$ near the boundary, $\Phi^A\simeq\phi^A(x)+ z^2\phi_{(2)}^A+\dots$, along with the target space metric, and choose Dirichlet boundary condition for the field $\Phi^A$. Restricting to a flat boundary, the asymptotic solution is
\begin{equation}\label{eq:solNLSM}
    \Phi^A\simeq\phi^A+ \frac{z^2}{4}\mathcal D^i\partial_i\phi^A+\mathcal O(z^4\log z)\,,
\end{equation}
where the Laplacian is now with respect to flat metric and the logarithmic term is again due to resonances in the equations of motion. Because of the radial powers in \cref{eq:actionNLSM}, using \cref{eq:solNLSM} we can expand the kinetic terms as
\begin{equation}
    \begin{aligned}
        \partial_z\Phi^A\,\partial_z\Phi^B
        + g^{ij}\partial_i\Phi^A\,\partial_j\Phi^B
        \simeq{}&
        \partial_i\varphi^A\,\partial^i\varphi^B +\frac{z^2}{4}
        \left(D^i\partial_i\varphi^A\right)
        \left(D^j\partial_j\varphi^B\right)\\
        & +\frac{z^2}{4}
        \partial_i\!\left(D^j\partial_j\varphi^A\right)
        \partial^i\varphi^B
        +\frac{z^2}{4}
        \partial_i\varphi^A\,
        \partial^i\!\left(D^j\partial_j\varphi^B\right)\\
        &
        +O(z^4\log z)\,.
    \end{aligned}
\end{equation}
Plugging back in \cref{eq:actionNLSM}, using near boundary expansion of the target space metric, integration by parts, and integrating out the radial direction, we get the on-shell action
\begin{equation}
    S_{\rm NLSM}=\frac{R^3_{\rm AdS}}{g_5^2}\int d^4x\, \mathcal G_{AB}\left(\frac{1}{4\epsilon^2}\partial_i\phi^A\partial^i\phi^B + \frac{\log\epsilon}{8}(\mathcal D^i\partial_i\phi^A)(\mathcal D^j\partial_j\phi^B)+\mathcal{O}(\epsilon^2)\right)\,.
\end{equation}
Thus, the logarithmic divergence comes from the \(z^2\) field correction in \cref{eq:solNLSM}. Now we can analytically continue the partition function. Once again, the real part of the wavefunction is generated by the phase of the logarithm as explained in \cref{sec:setup} and reads
\begin{equation}
    \log|\Psi_{\rm dS}|^2=-\frac{\pi R_{\rm dS}^3}{8  g_5^2}\int d^4x \,\mathcal G_{AB}(\mathcal D^i\partial_i\phi^A)(\mathcal D^j\partial_j\phi^B)\,.
\end{equation}
We further include the contributions from a $3$d and $7$d bulk in \cref{tab:summary-gr-YM-NLSM}.

\begin{table}[h]
    \centering
    \renewcommand{\arraystretch}{1.3}
    \setcellgapes{4pt}
    \makegapedcells
   \begin{tabularx}{\textwidth} { 
  | >{\centering\arraybackslash}p{0.2\textwidth} | >{\centering\arraybackslash}X | }
       \hline
        Theory & $\log|\Psi_{\rm dS}|^2$  \\
        \hline
        $5$d gravity & \(\displaystyle
        -\frac{M_{\rm pl}^3 R_{\rm dS}^3}{2} \frac{\pi}{8} \int d^4 x \sqrt{\hat{g}} \, {\hat{W}_{ijkl}} {\hat{W}^{ijkl}} 
        \)
        \\
        \hline
        $7$d gravity &\(\displaystyle
        \begin{aligned}
            -\frac{\pi M_{\rm pl}^5 R_{\rm dS}^5}{384}
            \int d^6x\,\sqrt{\hat g}\,
            \Bigg[
            &12\,\hat W_{iklj}\hat W^{imnj}\hat W_{m}{}^{kl}{}_{n}
            +3\,\hat W_{ij}{}^{kl}\hat W_{kl}{}^{mn}\hat W_{mn}{}^{ij}
            \\
            &+\hat W_{iklm}
            \left(
            \delta^i{}_j\hat\Box
            -4\hat R^i{}_j
            +\frac{6}{5}\hat R\,\delta^i{}_j
            \right)
            \hat W^{jklm}
            \Bigg]
            \end{aligned}
        \)
        \\
        \hline
        $5$d Yang-Mills & \(\displaystyle
        -\frac{ R_{\rm dS}}{4 g_5^2} \pi \int d^4 x  \hat F^a_{ij}\hat F^{a\,ij}
        \)\\
        \hline
        $7$d Yang-Mills &\(\displaystyle
        -\frac{R_{\rm dS}^3}{4g_7^2} \pi \int d^6x \frac{1}{2}\left[ (\hat{D}^j \hat{F}^a_{ji}) (\hat{D}^k \hat{F}^{a\, i}_{k}) \right]
        \) \\
        \hline
        $3$d NLSM & \(\displaystyle
        -\frac{\pi R_{\rm dS}}{2g_3^2}\int d^2x \,\mathcal G_{AB}(\phi)\partial_i\phi^A\partial^i\phi^B \)\\
        \hline
        $5$d NLSM &\(\displaystyle
        -\frac{\pi R^3_{\rm dS}}{8g_5^2}\int d^4x \,\mathcal G_{AB}(\phi)(\mathcal D^i\partial_i\phi^A)(\mathcal D^j\partial_j\phi^B)\) \\
        \hline
       $7$d NLSM  &
        \(\displaystyle
        \begin{aligned}
            -\frac{\pi R^5_{\rm dS}}{128 g_7^2}\int\! d^6x \Big[&\mathcal{D}_i(\mathcal{D}^j\partial_j\phi^A)\mathcal{D}^i(\mathcal{D}^k\partial_k\phi^B)\\
             &-R_{ABCD}(\mathcal{D}^j\partial_j\phi^A) \partial_i \phi^B (\mathcal{D}^k\partial_k\phi^C)\partial^i\phi^D\Big]
        \end{aligned}
        \) \\
        \hline
    \end{tabularx}
    \caption{Logarithm of the dS measure for gravity, Yang-Mills theory, and NLSM in different dimensions. In $7$d Yang-Mills, the factor of $\frac12$ in the integrand does not arise from a difference in normalization: the action is normalized according to the same convention as in $5$d. In $7$d NLSM, $R_{ABCD}$ is the Riemann tensor of the target space metric $\mathcal{G}_{AB}$.}
    \label{tab:summary-gr-YM-NLSM}
\end{table}

\section{Lorentzian Checks and Cosmological CPT}
\label{sec:direct_check_and_origin}

To illustrate these mechanisms purely on the dS side, we begin with an independent, explicit check of the local Yang-Mills probability measure in dS$_5$ using the Lorentzian wavefunction formalism. Specifically, we evaluate the two- and three-point wavefunction coefficients directly in the flat slicing,
\begin{equation}
    ds^2 = \frac{R_{\text{dS}}^2}{\eta^2} \left( -d\eta^2 + d\mathbf{x}^2 \right)\,, \qquad -\infty < \eta < 0\,,
\end{equation}
working in temporal gauge $A_\eta^a = 0$. The probability measure is determined by the imaginary part of the on-shell action, \cref{eq:log-measure}, which we compute order-by-order in the boundary field $\hat A$. We further present a partial explanation of the results using the cosmological CPT theorem \cite{Goodhew:2024eup}

\subsection{Quadratic Yang-Mills Coefficient}

The quadratic part of the action for the transverse spatial components ($k_i A_i^a = 0$) reads
\begin{equation}
    S_{\rm YM}\supset \frac{R_{\text{dS}}}{2g_5^2} \int \frac{d\eta}{(-\eta)}\,d^4x \left[ A_i^{a\prime} A_i^{a\prime} - \frac{1}{2} f_{ij}^a f_{ij}^a \right]\,,
\end{equation}
where $f_{ij} = \partial_i A_j - \partial_j A_i$. The linearized equation of motion, $A_i^{a\prime\prime} - \frac{1}{\eta}A_i^{a\prime} + k^2 A_i^a = 0$, is solved by the Bunch-Davies bulk-to-boundary propagator $A_i^a(\eta, \mathbf{k}) = K_k(\eta)\hat A_i^a(\mathbf{k})$, where
\begin{equation}
    K_k(\eta) = -\frac{i\pi}{2} (-k\eta) H_1^{(2)}(-k\eta)\,,
\end{equation}
where $H_1^{(2)}$ is the Hankel function of second kind of order 1. Using the small-argument expansion of the Hankel function,
\begin{equation}
    K_k(\eta) = 1 - \frac{k^2\eta^2}{2} \left[ \log\left(\frac{-k\eta}{2}\right) + \gamma_E - \frac{1}{2} + \frac{i\pi}{2} \right] + \mathcal{O}(\eta^4 \log\eta)\,,
\end{equation}
the quadratic part of the on-shell action reduces to the boundary term at late times ($\eta_* \to 0^-$):
\begin{equation}
    \begin{aligned}
        S_{\text{os}} &\supset
    -\frac{R_{\rm dS}}{2g_5^2\eta_*}
    \int \frac{d^4 k }{(2\pi)^4}
    A_i^a(\eta_*,\mathbf{k})
    A_i^{a\prime}(\eta_*,-\mathbf{k}) \\
    & = \frac{R_{\text{dS}}}{2g_5^2} \int \frac{d^4 k }{(2\pi)^4} k^2 \hat A_i^a(\mathbf{k}) \hat A_i^a(-\mathbf{k}) \left[ \log\left(\frac{-k\eta_*}{2}\right) + \gamma_E + \frac{i\pi}{2} \right]\,.
    \end{aligned}
\end{equation}
Extracting the imaginary part yields a purely local contribution:
\begin{equation}
    \operatorname{Im}S_{\text{os}} \supset \frac{\pi R_{\text{dS}}}{4g_5^2} \int \frac{d^4 k }{(2\pi)^4} k^2 \hat A_i^a(\mathbf{k}) \hat A_i^a(-\mathbf{k}) = \frac{\pi R_{\text{dS}}}{8g_5^2} \int d^4x\,\hat f_{ij}^a \hat f_{ij}^a\,,
\end{equation}
reproducing the expected local quadratic measure of \cref{eq:YM5_dS}.

\subsection{Cubic Yang-Mills Coefficient}
At cubic order, the interaction is governed by
\begin{equation}
    S_{\rm YM} \supset -\frac{R_{\text{dS}}}{2g_5^2} \int \frac{d\eta}{(-\eta)}\,d^4x\, f^{abc} f_{ij}^a A_i^b A_j^c\,.
\end{equation}
Evaluating this on the free solutions gives the cubic part of the on-shell action
\begin{equation}
    S_{\text{os}} \supset -\frac{R_{\text{dS}}}{2g_5^2} \int \frac{d^4 k_1 }{(2\pi)^4} \frac{d^4 k_2 }{(2\pi)^4} \frac{d^4 k_3 }{(2\pi)^4} (2\pi)^4 \delta^{(4)}\!\left(\sum_{a=1}^3 \mathbf{k}_a\right) \mathcal{V}_3 \, \mathcal{I}_3(k_1, k_2, k_3)\,,
\end{equation}
where $\mathcal{V}_3$ is the momentum-space cubic tensor
\begin{equation}
    \mathcal{V}_3
    =
    if^{abc}
    \left(
    k_{1i}\hat A_j^a(\mathbf{k}_1)
    -
    k_{1j}\hat A_i^a(\mathbf{k}_1)
    \right)
    \hat A_i^b(\mathbf{k}_2)
    \hat A_j^c(\mathbf{k}_3)\,,
\end{equation}
and the time-dependent integral is defined over the Bunch-Davies contour ($t = -\eta > 0$)
\begin{equation}
    \mathcal{I}_3(k_1, k_2, k_3) = \int_0^\infty \frac{dt}{t} \prod_{a=1}^3 \left[ -\frac{i\pi}{2} k_a t H_1^{(2)}(k_a t) \right]\,.
\end{equation}
To compute the imaginary part without getting embroiled in fully nonlocal finite expressions, we map $\mathcal{I}_3$ to the Euclidean triple-$K$ integral \cite{Bzowski:2013sza,BMSmixed} using the identity $-i\pi x H_1^{(2)}(x)/2 = ix K_1(ix)$. This yields the analytic continuation of the triple-$K$ integral in its momentum arguments, namely
\begin{equation}
    \mathcal{I}_3(k_1, k_2, k_3) = I_{2\{1,1,1\}}(ik_1, ik_2, ik_3)\,.
\end{equation}
The integral $I_{2\{1,1,1\}}$ possesses a logarithmic divergence at its lower endpoint ($z \to 0$), which we regulate by shifting the Bessel indices to $1+\varepsilon$. Then the divergent part of the regulated integral exhibits a simple pole \cite{BMSmixed}:
\begin{equation}
    I_{2\{1+\varepsilon, 1+\varepsilon, 1+\varepsilon\}}^{\text{div}}(k_a) = -\frac{1}{3\varepsilon} + F_{\text{real}}(k_a) + \mathcal{O}(\varepsilon)\,.
\end{equation}
The crucial mechanism for generating the imaginary part lies in the homogeneity property of the triple-$K$ integral under scaling, that is
\begin{equation}
    I_{2\{1+\varepsilon, 1+\varepsilon, 1+\varepsilon\}}(\lambda k_a) = \lambda^{3\varepsilon} I_{2\{1+\varepsilon, 1+\varepsilon, 1+\varepsilon\}}(k_a)\,.
\end{equation}
For the Lorentzian Bunch-Davies contour, we analytically continue the momenta by setting $\lambda = i$, giving
\begin{equation}
    I_{2\{1+\varepsilon, 1+\varepsilon, 1+\varepsilon\}}(ik_a) = e^{3i\pi\varepsilon/2} \left[ -\frac{1}{3\varepsilon} + F_{\text{real}}(k_a) + \mathcal{O}(\varepsilon) \right]\,.
\end{equation}
Expanding the exponential factor for small $\varepsilon$, the cross-term between the phase expansion and the simple pole produces a finite, universal imaginary contribution:
\begin{equation}
    \left( \frac{3i\pi}{2}\varepsilon \right) \left( -\frac{1}{3\varepsilon} \right) = -\frac{i\pi}{2}\,.
\end{equation}
Thus, the imaginary part of the integral originates exclusively from the log divergence pole:
\begin{equation}
    \operatorname{Im}\mathcal{I}_3(k_1, k_2, k_3) = -\frac{\pi}{2}\,.
\end{equation}
Because the pole is strictly independent of external momenta $k_a$, all non-trivial momentum-dependent pieces are relegated to the real part (the phase of the wavefunction), while the imaginary part remains entirely local.

Substituting this back into the cubic on-shell action yields the exact local measure contribution
\begin{equation}
    \operatorname{Im}S_{\text{os}}  \supset \frac{\pi R_{\text{dS}}}{4g_5^2} \int d^4x\, f^{abc} \hat f_{ij}^a \hat A_i^b \hat A_j^c\,,
\end{equation}
matching  the prediction from the EAdS analytic continuation through cubic order in \cref{eq:YM5_dS}.

\subsection{CPT Constraints and Locality}
\label{sec:origin}
The origin of the local measure in low-point Yang-Mills theory is related to the cosmological CPT theorem \cite{Goodhew:2024eup}. Specifically, below we demonstrate that the wavefunction coefficients of Yang-Mills theory in dS$_{d+1}$ do not contribute to the probability measure unless they exhibit logarithmic divergences. The same result goes through for NLSM and GR. This provides a purely de Sitter-based derivation of the phenomenon.

The wavefunction coefficients are defined through the perturbative expansion of the logarithm of the late-time wavefunction,
\begin{equation}
    \begin{aligned}
        \log\Psi[\hat A] & = \sum_{L=0}^{\infty}\sum_{n=2}^{\infty} \mathcal{W}^{(L)}_n [\hat{A}] \\
        & = \sum_{L=0}^{\infty}\sum_{n=2}^{\infty}
        \frac{1}{n!} \int \left[\prod_{a=1}^n \frac{d^dk_a}{(2\pi)^d} \hat A(\mathbf{k}_a) \right] (2\pi)^d\delta^{(d)} \left(\sum_{a=1}^n\mathbf{k}_a\right) T_n(\mathbf{k}_a) \hat\psi_n^{(L)}(\mathbf{k}_1, \ldots, \mathbf{k}_n),
    \end{aligned}
\end{equation}
where $L$ denotes the loop order, $T_n(\mathbf{k}_a)$ is the momentum-space tensor structure, and field indices and their contractions are left implicit. The cosmological CPT theorem \cite{Goodhew:2024eup} dictates that \textit{finite} wavefunction coefficients satisfy the reality-like condition 
\begin{equation}
    \left[\hat\psi_n^{(L)}\right]^*
    =
    e^{i\pi \left((d+1)(L-1) + dn - \sum_{a=1}^n\Delta_a\right)}
    \hat\psi_n^{(L)}\,,
\end{equation}
where $\Delta_a$ is the conformal dimension of the corresponding boundary operator. For a massless gauge field, the conformal dimension of the boundary operator is $\Delta_{\hat A} = d-1$. Therefore, at tree level ($L=0$), it reduces to 
\begin{equation}
    \left[\hat\psi_n^{(0)}\right]^*
    =
    e^{i\pi \left(-(d+1)+dn - n(d-1)\right)}
    \hat\psi_n^{(0)} 
    = 
    (-1)^{(-d-1+n)}
    \hat\psi_n^{(0)}\,.
\end{equation}
Note that this relation alone does not imply that the corresponding wavefunction terms are purely imaginary, as one must account for the parity-induced sign changes of the momentum-space tensor structures. To track these signs, consider the Yang-Mills action in temporal gauge, $A_\eta = 0$. Pure Yang-Mills theory features two types of interaction vertices: the cubic vertex, which contains a single spatial derivative,
\begin{equation}
    \int f^{abc}
    (\partial_iA_j^a-\partial_jA_i^a)
    A_i^bA_j^c\,,
\end{equation}
and the quartic vertex, which contains no spatial derivatives,
\begin{equation}
    \int f^{abc}f^{ade}
    A_i^bA_j^cA_i^dA_j^e\,.
\end{equation}
Under simultaneous spatial parity inversion, $\mathbf{k}_a \to -\mathbf{k}_a$, every cubic vertex contributes a factor of $-1$, whereas the quartic vertex does not. The propagators and transverse projectors are even under this transformation.

For a connected tree diagram with $V_3$ cubic vertices, $V_4$ quartic vertices, $I$ internal lines, and $n$ external legs, the tree topology implies
\begin{equation}
    I = V_3 + V_4 - 1\,.
\end{equation}
Counting the endpoints of all propagators yields
\begin{equation}
    3V_3 + 4V_4 = 2I + n\,.
\end{equation}
Substituting $I$ in the relation above gives the topological constraint
\begin{equation}
    n = V_3 + 2V_4 + 2\,,
\end{equation}
which implies $V_3 \equiv n \pmod 2\,$. 

For a real gauge field in position space, the Fourier components satisfy $\hat A_i^a(\mathbf{k})^* = \hat A_i^a(-\mathbf{k})$. Taking the complex conjugate of $\mathcal W_n$ and changing integration variables via $\mathbf{k}_a \to -\mathbf{k}_a$ leads to
\begin{equation}
    \mathcal W_n^*
    =
    (-1)^{n}
    \int \left[\prod_{a=1}^n \frac{d^dk_a}{(2\pi)^d} \hat A(\mathbf{k}_a) \right] (2\pi)^d\delta^{(d)} \left(\sum_{a=1}^n\mathbf{k}_a\right) T_n(\mathbf{k}_a) \left[\hat\psi_n^{(0)} \right]^{*}(\mathbf{k}_1, \ldots, \mathbf{k}_n)\,.
\end{equation}
Applying the CPT constraint $\hat\psi_n^* = (-1)^{n-d-1} \hat\psi_n$ then yields
\begin{equation}
    \mathcal W_n^*
    =
    (-1)^{n} (-1)^{n-d-1} \mathcal W_n = (-1)^{-d-1}\mathcal{W}_n\,,
\end{equation}
indicating that finite tree-level contributions are purely imaginary for odd $d+1$.

The non-triviality of the dS$_{d+1}$ probability measure arises from diagrams containing late-time logarithmic divergences, for which the simple finite-coefficient phase formula no longer applies. Consider a wavefunction coefficient with the following late-time asymptotic expansion
\begin{equation}
    \psi_n(-\eta;\mathbf{k}_a)
    = C_n(\mathbf{k}_a)\log(-\mu \eta) + B_n(\mathbf{k}_a)\,,
\end{equation}
where $C_n$ is the coefficient of the logarithmic divergence and $B_n$ denotes its finite counterpart. The cosmological CPT theorem implies that
\begin{equation}
    C_n^* = (-1)^{n+d+1} C_n\,,
\end{equation}
whereas the finite piece transforms as
\begin{equation}
    B_n^* = (-1)^{n+d+1} B_n - i\pi (-1)^{n+d+1} C_n\,.
\end{equation}
Accounting for the parity of the momentum structures, it follows that the non-vanishing contributions to the probability measure stem solely from the logarithmically divergent parts of the wavefunction coefficients. 

We can now relate the preceding argument to the explicit two- and three-point wavefunction calculations presented above. In both cases, the non-vanishing contributions to the probability measure are accompanied by logarithmic late-time divergences. Indeed, the quadratic on-shell action contains a term proportional to $\log(-\eta_*)\int \hat f^2$, while the cubic time integral $\mathcal I_3$ develops a logarithmic divergence as $\eta_*\to 0^-$. These observations are consistent with the cosmological CPT argument outlined above, according to which logarithmically divergent wavefunction coefficients evade the homogeneous reality constraint and can consequently give rise to nontrivial contributions to the probability measure.

To establish that the higher-point wavefunction coefficients do not contribute to the measure, however, the CPT argument alone is insufficient. One must additionally show that the coefficients with $n\geq 5$ do not develop late-time logarithmic divergences, so that the ordinary CPT reality constraint applies without the logarithmic exception. Finiteness of these coefficients can be deduced from the asymptotic solution of the equation of motion that is the Lorentzian analogue of \cref{eq:YM-final-expansion}.

We have therefore obtained a purely Lorentzian explanation of the structure of the tree-level Yang-Mills measure in $5$d dS: the CPT reality constraint eliminates the finite nonlocal contributions, while the only surviving contributions arise from logarithmic divergences whose coefficients are local functionals of the boundary gauge field.

\section{Concluding Thoughts and Future Directions}
\label{sec:conclusion}

In this paper, we have shown that a broad class of theories of interest in odd-dimensional de Sitter space have local and remarkably simple measures. More precisely, for Einstein gravity, Yang-Mills theory, and NLSM, nonlocal parts of the wavefunction reside entirely in its phase at the semiclassical level. This simplification follows from the logarithmic divergences of the corresponding Euclidean AdS on-shell actions, which, upon analytic continuation, entirely determine the de Sitter probability measure. We emphasize that the oddness of the bulk dimension is essential to our result. Indeed, massless fields in dS$_4$ exhibit nonlocal probability kernels already at quadratic order. It would also be interesting to understand this dimensional dependence directly from the Hankel expansions of bulk propagators, potentially identifying which higher-point interactions generate logarithmic divergences.

The cosmological CPT theorem offers a complementary perspective on the same mechanism. Suitably applied, through its reality-like conditions, it implies that finite tree-level wavefunction coefficients in the theories considered here contribute only to the phase of the wavefunction in odd-dimensional de Sitter space. Nevertheless, logarithmically divergent coefficients evade the corresponding reality condition. The main limitation of this approach is that it does not determine which coefficients develop such divergences, requiring them to be examined individually. In the EAdS derivation, this information is encoded in the local structure of $S_{\log}^{\rm EAdS}$. It is worth noting that the coefficient $S_{\log}^{\rm EAdS}$ is closely related to the holographic conformal anomaly. Our result therefore relates the computation of late-time correlation functions in de Sitter space to that of correlation functions in certain classically conformal local Euclidean field theories. This discussion should, however, be understood within its perturbative regime of validity. In particular, more general gravitational saddles need not admit a real EAdS continuation, in which case the finite part of the continued on-shell action can acquire an imaginary component and give additional, potentially nonlocal, contributions to the probability measure \cite{Cotler:2026flq, Cotler:2026xke}.

At the same time, the locality of the probability measure should not be confused with the locality of late-time correlation functions. Equal-time expectation values are obtained by integrating field insertions against $|\Psi|^2$,
\begin{equation}
    \left\langle \hat\Phi_1\cdots\hat\Phi_n \right\rangle
    =
    \frac{
    \displaystyle\int\mathcal D\hat\Phi\,
    \hat\Phi_1\cdots\hat\Phi_n
    |\Psi[\hat\Phi]|^2
    }{
    \displaystyle\int\mathcal D\hat\Phi\,
    |\Psi[\hat\Phi]|^2
    }\,.
\end{equation}
Even for a local probability action, the propagator is typically nonlocal. Consequently, exchange diagrams built from local probability vertices give rise to nonlocal connected correlators.

The dS$_5$ Yang-Mills example illustrates an additional simplification. Since the logarithm of its probability measure is proportional to $\int\widehat F^2$, the elementary probability vertices terminate at quartic order. Yet, connected correlators of arbitrarily high multiplicity are generally nonzero and can be constructed recursively by joining the cubic and quartic probability vertices with propagators.

It is also natural to ask whether this simplicity survives the inclusion of additional interactions. The EAdS argument suggests that locality is not restricted to the two-derivative theories studied here. One can consider higher-curvature gravitational interactions, higher powers of the Yang-Mills field strength, shift-symmetric scalar interactions, and higher-derivative interactions of general $p$-form gauge fields. The argument continues to hold, generically because contractions with the bulk inverse metric yield a factor of $R_{\rm EAdS}^{d+1-2m}$, where $m$ counts the inverse metrics. In odd bulk dimensions, this remains an odd power of the radius, so the same analytic continuation applies. Hence, provided the near-boundary expansions of the fields remain analytic in the radial direction $z$, higher-derivative interactions may modify the local functional determining the probability measure without spoiling its locality.

Another possible generalization concerns semi-local measures, whose wavefunction coefficients have support only when subsets of the boundary insertion points coincide. For instance, a three-point kernel may contain a term of the form
\begin{equation}
    \mathcal K_3(x_1,x_2,x_3)
    \supset
    \delta^{(d)}(x_1-x_2)\,
    \mathcal F(x_1,x_3)\,,
\end{equation}
where $\mathcal F$ is generically nonlocal. Interacting scalar theories may provide examples of this structure. A generic $\phi^n$ potential can modify the leading asymptotic behavior of a massless scalar and invalidate the standard expansion around arbitrary finite Dirichlet boundary conditions. It would be interesting to investigate semi-local contributions from the perspective of cosmological CPT, particularly whether it constrains their possible forms and enables a broader classification of late-time probability measures.

The case of three-dimensional de Sitter space deserves separate considerations. Pure gravity has no local propagating degrees of freedom, and its 2d logarithmic functional is topological on a closed boundary. The leading order dynamics of the gauge fields is controlled by Chern-Simons theory, which again is topological and produces a wavefunction which is a pure phase.\footnote{For a recent study of Chern-Simons theory in AdS$_3$, see \cite{Mathys:2026pdq}.} Although the topological nature of gauge fields and gravity makes the three-dimensional case particularly simple, nontrivial results can still arise in scalar theories, as illustrated by the NLSM.

Our results further ask for a comparison with recent Grassmannian formulations of four-dimensional CFT correlators \cite{Bala:2026trw}. Since the five-dimensional measures studied here are obtained by analytic continuation from EAdS$_5$, these constructions may provide an independent way of examining the locality of the probability measure. In particular, it would be useful to understand how the logarithmic contributions associated with holographic anomalies are encoded in the Grassmannian framework.

An important open question is how locality is modified beyond tree level. Loops can introduce anomalous dimensions, additional logarithms, and nonlocal contributions that cosmological CPT alone does not exclude. More generally, understanding how far the simplification we have found extends, and whether it admits a more direct explanation in terms of the underlying structure of cosmological wavefunction coefficients, remains an open problem.

\acknowledgments 
We are grateful to Carlos Duaso Pueyo, Emine Şeyma Kutluk, Manuel Loparco, Juan Maldacena, Charlotte Sleight, and Massimo Taronna for helpful comments and discussions. GLP, NR and BT are supported by the ERC (NOTIMEFORCOSMO, 101126304), by Scuola Normale, and by INFN (IS GSS-Pi). GLP is also supported by the Italian Ministry of Universities and Research (MUR) under contract 20223ANFHR (PRIN2022). NR is further supported by the European Union’s Horizon Europe research and innovation programme under the Marie Skłodowska-Curie Actions – Global Postdoctoral Fellowship (ACROSS, 101267248).

\bibliographystyle{JHEP}
\bibliography{biblio}

\end{document}